\documentclass[prx,aps,twocolumn,longbibliography,superscriptaddress]{revtex4-2}

\usepackage[pdftex]{graphicx}
\usepackage{amsbsy,amssymb,amsmath,bm,mathtools}

\usepackage{xspace}
\usepackage{bm}
\usepackage{color}
 
\usepackage{url}
\usepackage[section]{placeins}
\usepackage[colorlinks=true,linkcolor=blue,citecolor=blue]{hyperref}

\usepackage{comment}
\usepackage{graphicx}
\usepackage{color}
\usepackage{epstopdf}

\begin{document} 
\let\vec\mathbf

\title{Learning Degenerate Manifolds of Frustrated Magnets with Boltzmann Machines}

\author{Ho Jang}
\affiliation{Department of Physics, University of Virginia, Charlottesville, VA 22904, USA}

\author{Jackson C. Glass}
\affiliation{Department of Physics, University of Virginia, Charlottesville, VA 22904, USA}

\author{Gia-Wei Chern}
\affiliation{Department of Physics, University of Virginia, Charlottesville, VA 22904, USA}

\graphicspath{{Figures/}}

\date{\today}

\begin{abstract}
We show that Restricted Boltzmann Machines (RBMs) provide a flexible generative framework for modeling spin configurations in disordered yet strongly correlated phases of frustrated magnets. As a benchmark, we first demonstrate that an RBM can learn the zero-temperature ground-state manifold of the one-dimensional ANNNI model at its multiphase point, accurately reproducing its characteristic oscillatory and exponentially decaying correlations. We then apply RBMs to kagome spin ice and show that they successfully learn the local ice rules and short-range correlations of the extensively degenerate ice-I manifold. Correlation functions computed from RBM-generated configurations closely match those from direct Monte Carlo simulations. For the partially ordered ice-II phase---featuring long-range charge order and broken time-reversal symmetry---accurate modeling requires RBMs with uniform-sign bias fields, mirroring the underlying symmetry breaking. These results highlight the utility of RBMs as generative models for learning constrained and highly frustrated magnetic states.
\end{abstract}


\maketitle

\section{introduction}

\label{sec:intro}

Machine learning (ML) has seen rapid adoption across physics and chemistry in recent years~\cite{carleo19,sarma19,bedolla21,schwartz21,boehnlein22,karagiorgi22}. Supervised approaches now play a central role in predicting materials properties,  constructing phase diagrams, and classifying quantum states. A notable success is the development of ML-based interatomic potentials, which now underpin {\em ab initio} molecular dynamics (MD) simulations~\cite{behler07,bartok10,li15,shapeev16,behler16,botu17,smith17,zhang18,deringer19,mcgibbon17,suwa19,chmiela17,chmiela18,sauceda20,unke21} and a growing range of condensed-matter applications~\cite{zhang20,zhang21,zhang22,zhang22b,zhang23,cheng23,cheng23b,Ghosh24,Fan24,Ma19,Liu22,Tian23}.  In parallel, unsupervised learning has become an indispensable tool for uncovering structure in unlabeled data: methods ranging from PCA to modern generative models such as variational autoencoders (VAEs) extract dominant modes, reveal latent variables, and capture complex many-body correlations. Together, these ML techniques have enabled the identification of new phases of matter, the detection of emergent features in quantum systems, and the discovery of hidden organizing principles in large-scale simulation and experimental datasets.

A prototypical unsupervised generative model is the Restricted Boltzmann Machine (RBM)~\cite{hinton86,smolensky87,hinton89,hinton02}. An RBM is designed to learn the probability distribution underlying a dataset by reconstructing inputs from a set of latent (hidden) variables. Its architecture consists of a visible layer representing the data and a hidden layer encoding latent features, with full connectivity between layers but no intra-layer couplings---hence the ``restricted’’ bipartite structure. This design enables efficient evaluation of visible–hidden correlations and supports stochastic training procedures, such as contrastive divergence, that approximate maximum-likelihood learning. By capturing a compact probabilistic representation of the data, RBMs provide powerful tools for generative modeling, feature extraction, and dimensionality reduction.

As generative models, RBMs have shown a remarkable ability to reproduce equilibrium properties of many-body systems~\cite{torlai16,morningstar18,yevick21,decelle21,zhang25}. Trained on Monte Carlo configurations of the square-lattice Ising model, for instance, an RBM can learn the underlying equilibrium distribution and generate new samples that accurately reproduce heat-capacity, magnetization, and susceptibility. Their generative power has also inspired hybrid Monte Carlo schemes in which RBMs propose nonlocal, cluster-type updates that improve sampling efficiency~\cite{huang17,wang17,shen18,pilati19,puente20}. Beyond classical systems, RBMs have also been employed as expressive variational ans\"atze for quantum many-body states~\cite{carleo17,nomura17,torlai2018,melko19}. In this context, the RBM represents the complex wave function amplitudes in a compact latent space, enabling efficient variational optimization of ground states. These successes reflect a deep structural analogy between RBMs and equilibrium statistical mechanics: the RBM energy function closely mirrors a classical spin Hamiltonian. Consequently, RBMs serve both as practical computational tools and as conceptual bridges linking modern machine learning with statistical and quantum many-body physics.

In this work, we apply RBMs to a qualitatively different class of classical statistical-physics problems. Instead of focusing on ordered phases with relatively simple spatial correlations, we investigate whether an RBM can learn the highly correlated yet intrinsically disordered spin-liquid phases that arise in strongly frustrated magnets. In such systems, pairwise spin interactions cannot be simultaneously satisfied—whether due to special lattice geometries such as corner-sharing triangles or tetrahedra, or due to competing exchange interactions. This frustration leads to an extensively degenerate ground-state manifold, characterized by nontrivial local constraints and intricate correlation patterns that persist without conventional symmetry breaking.

Our central goal is to determine whether an RBM can faithfully approximate the underlying probability distribution of these spin-liquid ensembles, despite the absence of long-range order and the subtle constraints governing their low-energy manifold. More broadly, we seek to assess the capabilities and limitations of RBMs as generative models for frustrated, highly degenerate classical systems --- examining whether their compact latent representations can capture emergent constraints, gauge-like degrees of freedom, and long-range correlations characteristic of classical spin liquids.

As concrete case studies, we apply the RBM framework to two paradigmatic examples: the extensively degenerate ground-state manifold of the one-dimensional axial next-nearest-neighbor Ising (ANNNI) model~\cite{elliott61,fisher80,bak82,selke88}, and the correlated spin configurations of two-dimensional kagome spin-ice systems~\cite{wills02,moller09,chern11,chern12}. These systems provide complementary testbeds---one with one-dimensional, constraint-driven degeneracy, and the other with two-dimensional, Coulomb-phase-like correlations---allowing us to probe the robustness of RBM-based generative modeling across distinct forms of frustration.

The remainder of this paper is organized as follows. In Sec.~\ref{sec:RBM}, we briefly review the RBM architecture and training procedures relevant to probabilistic modeling of classical spin configurations. Section~\ref{sec:annni} presents our results for the ANNNI model, highlighting the RBM’s ability to learn constraint-dominated ensembles. Section~\ref{sec:kagome} then examines the kagome spin-ice system, focusing on the RBM's capacity to reproduce algebraic correlations and divergence-free conditions. We conclude in Sec.~\ref{sec:conclusion} with a summary of our findings and a discussion of broader implications for ML-based generative modeling of frustrated and highly degenerate statistical systems.

\section{Restricted Boltzmann Machines}

\label{sec:RBM}

Restricted Boltzmann machines (RBMs) are energy-based generative models defined on a bipartite graph comprising visible and hidden units. Introduced as a tractable specialization of the fully connected Boltzmann machine, RBMs eliminate all intra-layer couplings, thereby yielding factorized conditional distributions and enabling efficient Gibbs sampling and approximate maximum-likelihood training. This architectural simplification makes RBMs well suited for modeling high-dimensional probability distributions and has contributed to their widespread use in machine-learning applications since the mid-2000s.

\begin{figure}
\centering
\includegraphics[width=0.99\columnwidth]{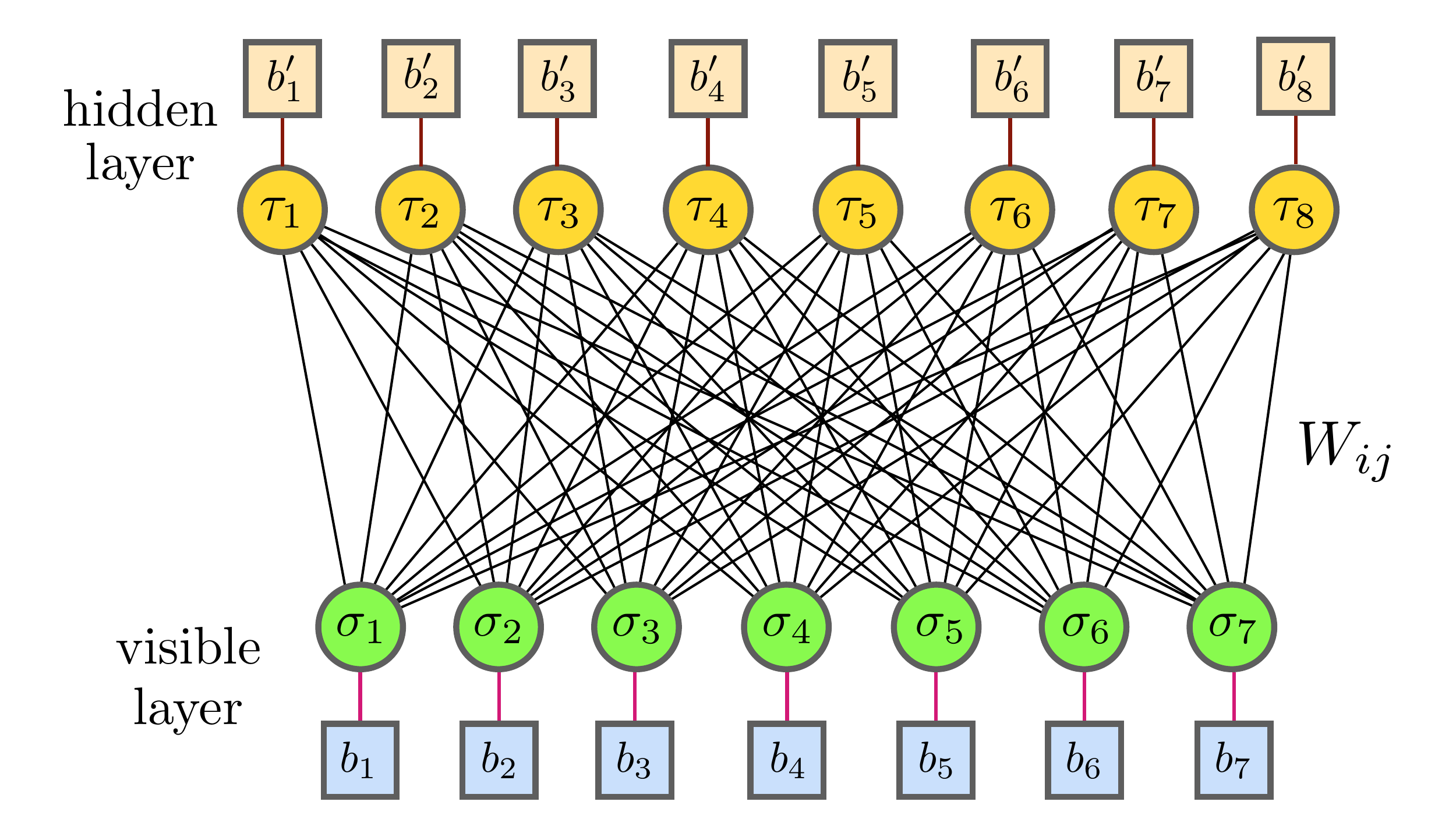}
\caption{Schematic diagram of a restricted Boltzmann machine (RBM) in an Ising-spin representation. The visible spins $\bm{\sigma}$ (green) are coupled to the hidden spins $\bm{\tau}$ (yellow) through the weight matrix $\bm W$. The local fields $\mathbf{b}$ and $\mathbf{b}'$ appearing in the RBM Hamiltonian are indicated by external boxes connected to the visible and hidden layers, respectively.}
\label{fig:RBM}
\end{figure}

From a physics perspective, an RBM may be viewed as a bipartite spin-glass model defined on two sets of Ising variables, $\bm\sigma = (\sigma_1, \sigma_2, \cdots, \sigma_N)$ and $\bm\tau = (\tau_1, \tau_2, \cdots, \tau_M)$; see Fig.~\ref{fig:RBM}. Here $N$, $M$ are the number of spins (neurons) in the visible and hidden layers, respectively. Every visible spin couples to every hidden spin through the edges of the bipartite graph. The associated coupling strengths are collected in a weight matrix $\mathbf W$ with vanishing diagonal, where $W_{ij}$ parameterizes the interaction between $\sigma_i$ and $\tau_j$. Two sets of local fields, $\bm b = (b_1, b_2, \cdots, b_N)$ and $\bm b' = (b'_1, b'_2, \cdots, b'_M)$, act on the visible and hidden layers, respectively. The joint distribution realized by the RBM is an ``equilibrium" Boltzmann distribution,\begin{eqnarray}
 	\pi_{\bm\theta}(\bm\sigma, \bm\tau) = \frac{1}{\mathcal{Z}_{\bm\theta}} e^{-E_{\bm \theta}(\bm\sigma, \bm\tau)},
\end{eqnarray}
where $\bm\theta = \{W_{ij}, b_i, b'_j \}$ denotes the full set of parameters, $\mathcal{Z}_{\bm\theta}$ is the partition function, and the effective energy takes the form
\begin{eqnarray}
	E_{\bm\theta}(\bm\sigma, \bm\tau) = -\sum_{i=1}^N \sum_{j = 1}^M W_{ij} \sigma_i \tau_j - \sum_{i=1}^N b_i \,\sigma_i - \sum_{j = 1}^M b'_j\, \tau_j. \nonumber \\
\end{eqnarray}
The bipartite structure also enables efficient nonlocal dynamical updates. Because the state of each node is sampled from a nonlinear function of its inputs (its ``activation"), and because nodes within the same layer are conditionally independent, one may update all visible units simultaneously given the hidden layer, and vice versa. This layer-wise sampling can be implemented using fast linear-algebra operations. In particular, for any fixed parameter set $\bm\theta$, the conditional distributions $p_{\bm\theta}(\bm\sigma \,|\, \bm\tau)$ and $p_{\bm\theta}(\bm\tau \,|\, \bm\sigma)$ factorize over sites and are obtained directly via Bayes' rule:
\begin{eqnarray}
	p_{\bm\theta}(\bm\sigma \big| \bm\tau) = \prod_{i=1}^N p_{\bm\theta}(\sigma_i \big| \bm\tau),\quad
	p_{\bm\theta}(\bm\tau \big| \bm\sigma) = \prod_{j=1}^M p_{\bm\theta}(\tau_j \big| \bm\sigma)
\end{eqnarray}
with the single-site probabilities
\begin{eqnarray}
	& & p_{\bm\theta}(\sigma_i = 1 \big| \bm\tau) = {\rm sigm}\Bigl(\sum_j W_{ij} \tau_j + b_i\Bigr), \nonumber \\
	& & p_{\bm\theta}(\tau_j = 1 \big| \bm\tau) = {\rm sigm}\Bigl(\sum_i W_{ij} \sigma_i + b'_j\Bigr)
\end{eqnarray}
where ${\rm sigm}(x) = (1+e^{-x})^{-1}$ denotes the logistic (sigmoid) activation function.

As a generative model, the RBM represents a target probability distribution through the marginal obtained by integrating out the hidden spins:
\begin{eqnarray}
	\label{eq:p_marginal}
	\pi_{\bm\theta}(\bm\sigma) = \sum_{\bm \tau} \pi_{\bm\theta}(\bm\sigma, \bm\tau) = \frac{1}{\mathcal{Z}_{\bm \theta}} \sum_{\bm \tau} e^{-E_{\bm\theta}(\bm\sigma, \bm\tau)}.
\end{eqnarray}
Integrating out the hidden layer induces nontrivial, effectively higher-order interactions among the visible spins. Owing to the bipartite structure of the RBM, this marginalization can be performed analytically, yielding
\begin{eqnarray}
	\pi_{\bm\theta}(\bm \sigma) = \frac{1}{\mathcal{Z}_{\bm\theta}} e^{- E_{\bm\theta}(\bm\sigma)},
\end{eqnarray}
with an effective free energy of the visible spins:
\begin{eqnarray}
	\label{eq:E_eff}
	E_{\bm\theta}(\bm\sigma) = -\sum_i b_i \sigma_i - \sum_j \ln\left(1 + e^{b'_j + \sum_i W_{ji} \sigma_i} \right)
\end{eqnarray}
The second term encapsulates the influence of the hidden spins and generates highly nonlinear couplings among the visible units, enabling the RBM to model complex correlated manifolds. Training the RBM amounts to adjusting $\bm\theta$ so that $\pi_{\bm\theta}(\bm\sigma)$ approximates a desired target distribution. Once trained, new samples are produced by alternating Gibbs updates of the visible and hidden layers, an efficient sampling strategy afforded by the bipartite architecture.

In the present context, the RBM is tasked with approximating the Boltzmann distribution of a classical spin Hamiltonian $\mathcal{H}_{\rm phys}(\bm\sigma)$,
\begin{eqnarray} \label{eq:phys_distribution}
	\pi_{\rm phys}(\bm\sigma) = \frac{e^{-\beta \mathcal{H}_{\rm phys}(\bm\sigma)}}{\mathcal{Z}_{\rm phys}},
\end{eqnarray}
where $\beta = 1/k_B T$ and $\mathcal{Z}_{\rm phys} = \sum_{\bm\sigma} e^{-\beta \mathcal{H}_{\rm phys}(\bm\sigma)}$. 
The training objective is to minimize the Kullback--Leibler divergence between $\pi_{\bm\theta}$ and $\pi_{\rm phys}$. The parameters evolve according to the stochastic gradient update
\begin{eqnarray}
	\label{eq:update}
	\bm\theta^{(n+1)} = \bm\theta^{(n)} + \eta \left[ \left\langle -\frac{\partial E_{\bm\theta}}{\partial \bm\theta} \right\rangle_{\rm data} 
	- \left\langle -\frac{\partial E_{\bm\theta}}{\partial \bm\theta} \right\rangle_{\rm model} \right], \quad
\end{eqnarray}
where the data average is computed by clamping the visible layer to a dataset sample $\tilde{\bm \sigma}$ and sampling the hidden layer via $p(\bm\tau|\bm\sigma=\tilde{\bm \sigma})$, while the model average is approximated using contrastive divergence (CD-$k$). Through this procedure, the RBM learns to generate spin configurations that faithfully capture the intricate low-temperature structure of frustrated magnetic systems. Additional implementation details are provided in Appendix~\ref{app:numerical-details}.

\section{ANNNI model}
\label{sec:annni}

The anisotropic next-nearest-neighbor Ising (ANNNI) model~\cite{elliott61,fisher80,bak82,selke88} is a canonical classical spin system for studying frustrated magnetism, modulated phases, and commensurate-incommensurate transitions. By supplementing the ferromagnetic Ising model with an antiferromagnetic next-nearest-neighbor interaction along a single spatial axis, the model introduces competition between couplings of different signs and ranges. This interaction-induced frustration---present even on bipartite lattices---stabilizes spatially modulated configurations that do not arise in the nearest-neighbor Ising model. Because the underlying symmetry remains purely Ising-like and geometric frustration is absent, the ANNNI model provides a clean theoretical setting for analyzing complex ordering phenomena generated solely by competing interactions. 

\begin{figure}
\centering
\includegraphics[width=0.99\columnwidth]{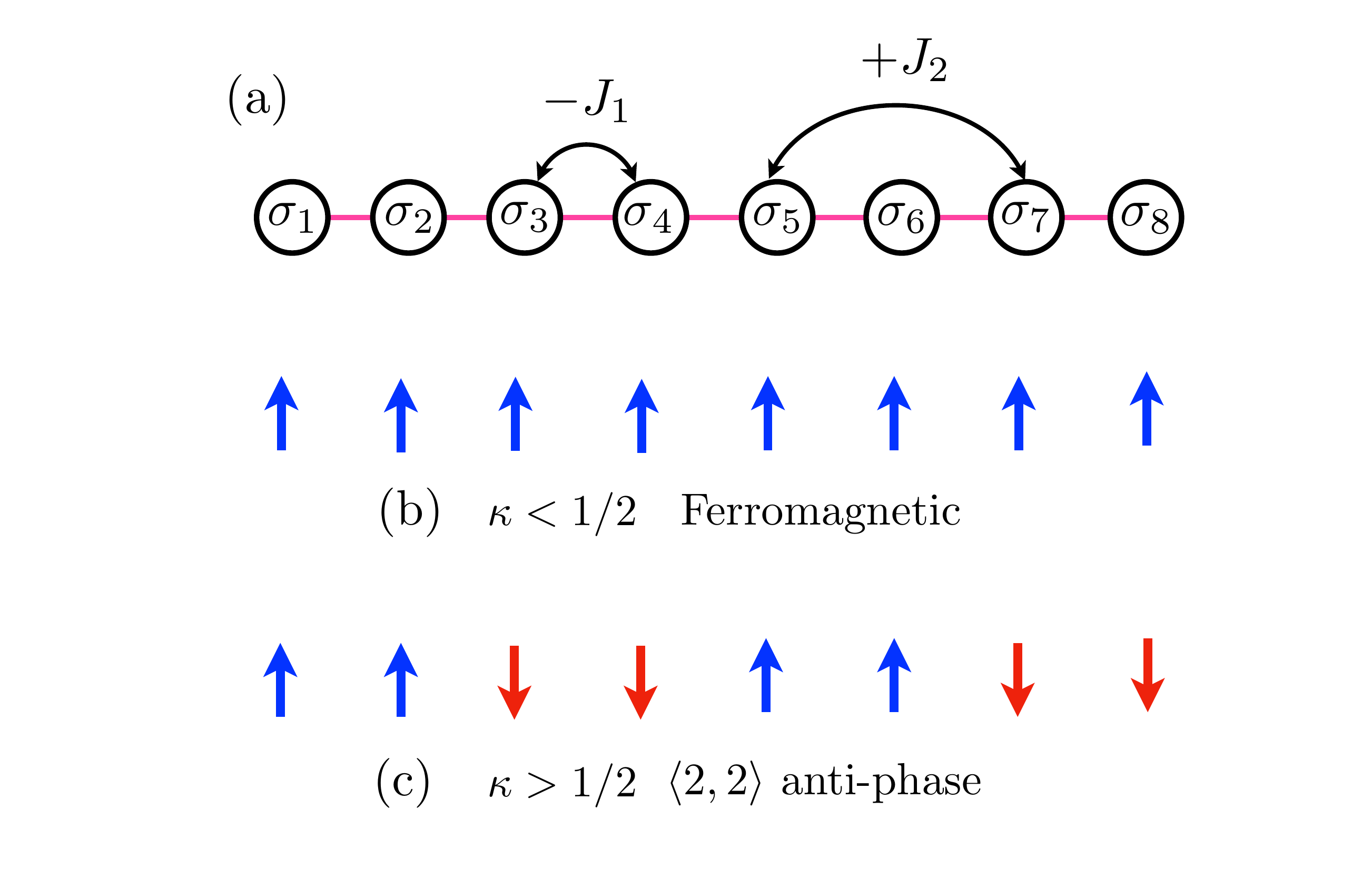}
\caption{(a) Competing interactions in the 1D ANNNI chain: ferromagnetic nearest-neighbor bonds $J_1$ and antiferromagnetic next-nearest-neighbor bonds $J_2$. (b) Ferromagnetic ground state for $\kappa = J_2 / J_1 < 1/2$, and (c) the anti-phase $\langle 2, 2\rangle$ state for $\kappa > 1/2$. }
\label{fig:annni-schematic}
\end{figure}

In higher dimensions, such competition produces a finite-temperature phase diagram containing an infinite hierarchy of commensurate modulated phases. Between the uniform ferromagnet and the period-2 antiphase, the model supports a dense sequence of ordered structures of the form $\uparrow^m \downarrow^n$, each characterized by a rational modulation wave vector. These states form a devil’s staircase: a fractal hierarchy of locked plateaus that proliferate near the $T=0$ boundary separating the uniform and period-2 phases, and expand into a characteristic fan-shaped region of commensurate order at finite temperature. At still higher temperatures, this staircase gives way to an incommensurate (floating) phase in which the modulation wave vector varies continuously.

As a warm-up example for the RBM application, we focus on the one-dimensional ANNNI chain with Hamiltonian
\begin{eqnarray}
	\mathcal{H}_{\rm ANNNI}(\bm \sigma) = -J_1 \sum_{i=1}^N \sigma_i \sigma_{i+1} + J_2 \sum_{i=1}^N \sigma_i \sigma_{i+2},  
\end{eqnarray}
where $J_1>0$ is the ferromagnetic nearest-neighbor interaction and $J_2>0$ is the antiferromagnetic next-nearest-neighbor coupling along the axial direction; see Fig.~\ref{fig:annni-schematic}(a) for a schematic of the model. Periodic boundary condition is assumed. Although the 1D model lacks true long-range order at any finite temperature, its zero-temperature phase diagram already captures the essential effects of frustration. For $\kappa = J_2/J_1 < 1/2$, the ground state is the uniform ferromagnetic configuration $\uparrow\uparrow\uparrow\cdots$, determined by the dominant nearest-neighbor term. For $\kappa > 1/2$, frustration stabilizes a modulated period-2 antiphase of the form $\uparrow\uparrow\downarrow\downarrow\cdots$. At the special point $\kappa = 1/2$, the domain-wall energy vanishes and the model develops a macroscopically large manifold of degenerate ground states: any spin sequence composed solely of domains of length one or two has identical energy density, reflecting the collapse of the effective tension between adjacent spin blocks.

Here we are interested in the special point $\kappa = 1/2$,  where the competition between $J_1$ and $J_2$ leads to extensive degeneracy. Although the $\kappa = 1/2$ at $T = 0$ is not a critical point in the sense of a divergent correlation length, it nevertheless exhibits highly nontrivial short-range correlations rooted in the macroscopic degeneracy. At $T = 0$ and $\kappa = 1/2$, any spin configuration with no three consecutive spins of the same sign is a ground state. The number of such configurations grows as $\mu^N$~\cite{redner81}, where $\mu = (1+\sqrt{5})/2$ is the golden ratio, so the residual entropy per spin is $S_0 = \ln \mu$.

In the natural ensemble in which all ground states are weighted equally, the longitudinal spin-spin correlator is short-ranged and oscillatory~\cite{dhar00}. A transfer-matrix analysis over the constrained ground-state manifold shows that, in the thermodynamic limit, $C(r) = A \cos(2 \pi r/3 + \phi) e^{-r/\xi}$ with a finite correlation length $\xi = 1/\ln(\frac{3+\sqrt{5}}{2}) \approx 1.04$, and the amplitudes/phases $A$ and $\phi$ that differ between even and odd sites.

\begin{figure}[b]
\centering
\includegraphics[width=0.99\columnwidth]{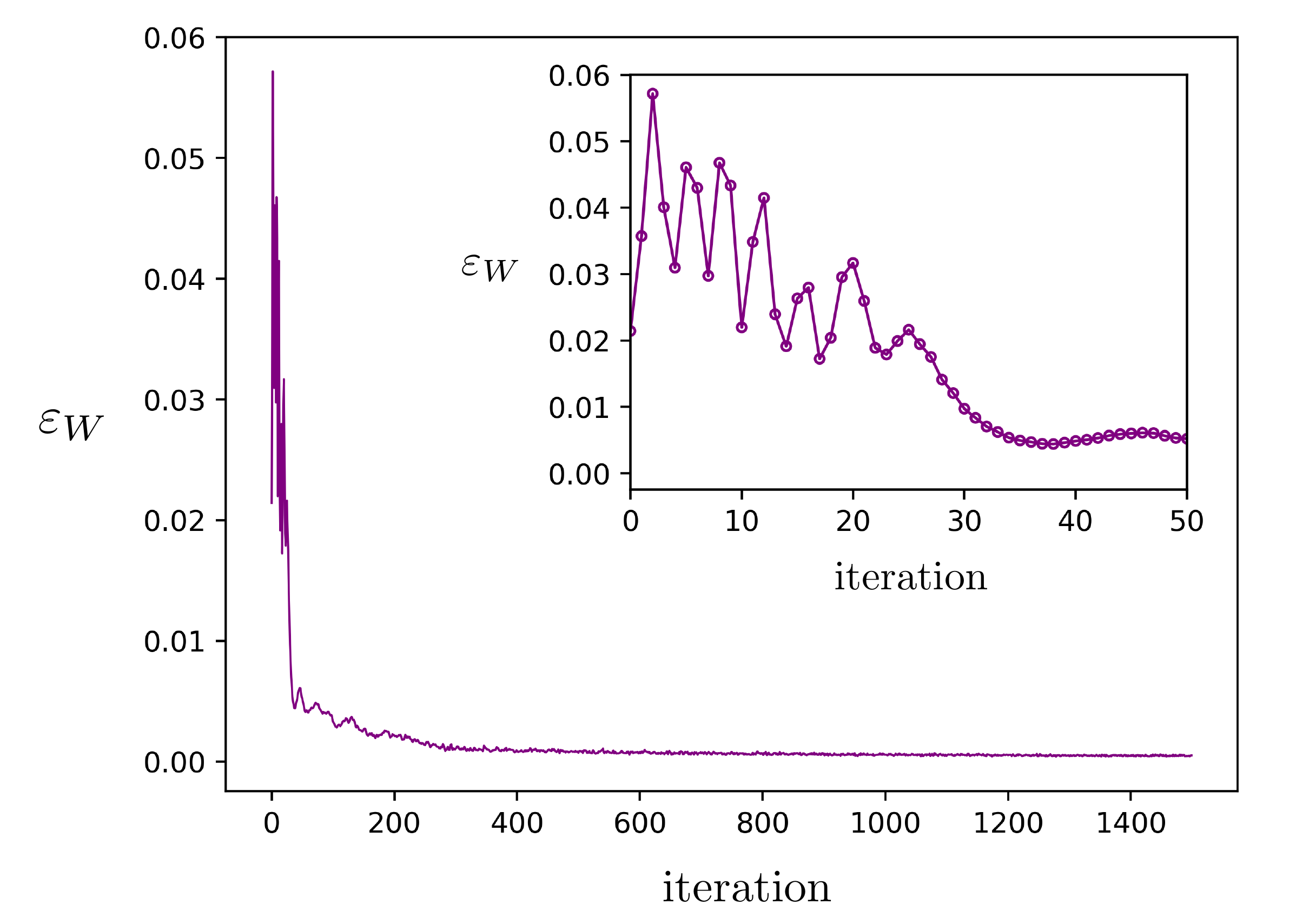}
\caption{Plot of $\varepsilon_W$ during training of the ice-II phase RBM. Early oscillatory behavior indicates rapid adjustment of model parameters, while decay of $\varepsilon_W$ at later iterations indicates convergence of the gradient descent procedure.}
\label{fig:training-annni}
\end{figure}

\begin{figure}
\centering
\includegraphics[width=0.99\columnwidth]{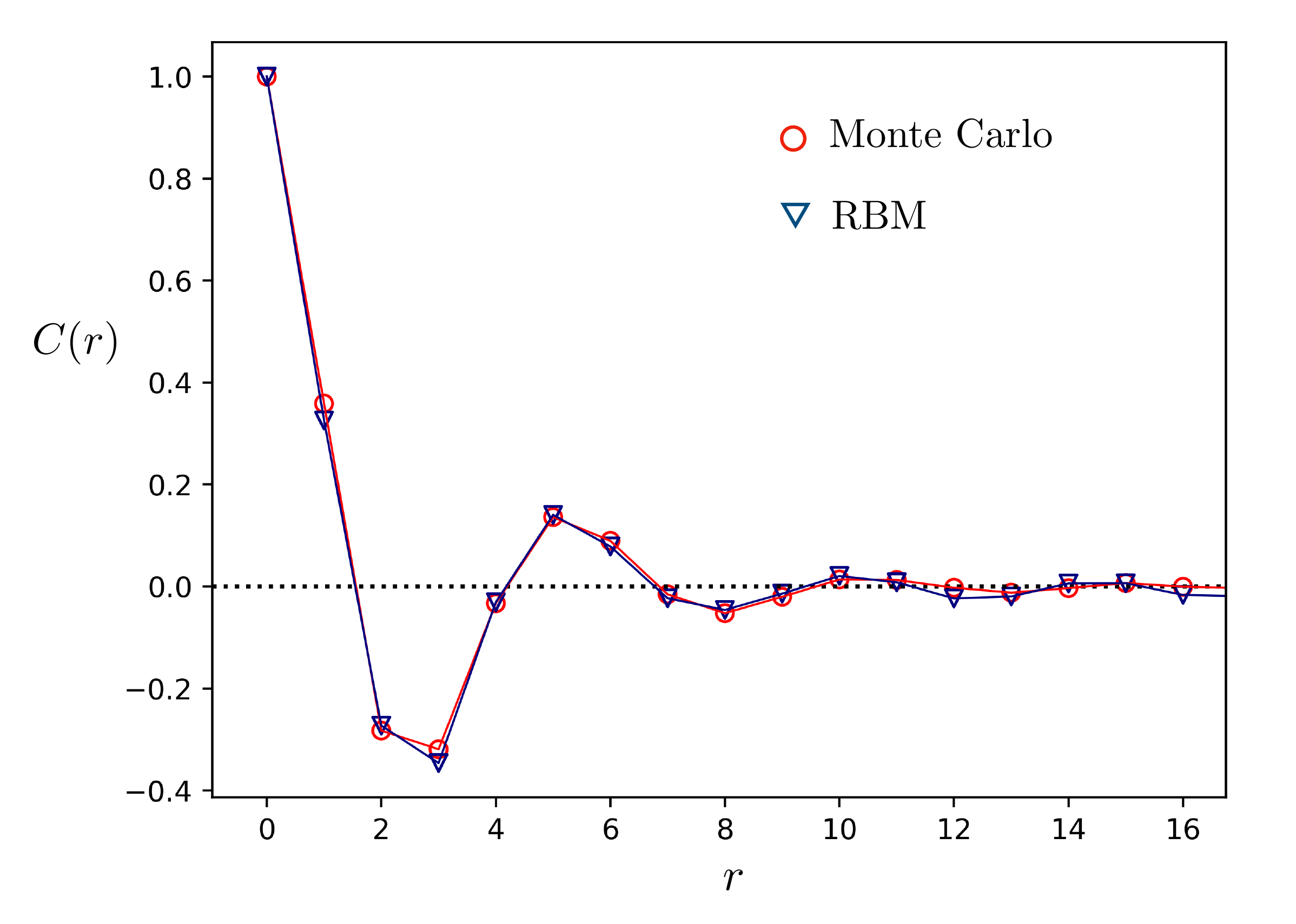}
\caption{Comparison of the spin-spin correlation function $C(r)$ obtained from Markov-chain Monte Carlo sampling and from RBM-generated configurations for the ANNNI chain at the multiphase point $\kappa = 1/2$ and $T = 0$.}
\label{fig:corr-annni}
\end{figure} 

Since we aim to train an RBM at the multiphase point $\kappa = 1/2$ and $T = 0$, the corresponding target distribution is simply uniform over the degenerate ground-state manifold. More explicitly, $\pi_{\rm ANNNI}(\bm \sigma; \, \kappa = 1/2, T = 0) = 1/\mathcal{Z}$ for any spin configuration containing no three consecutive spins of the same sign, where $\mathcal{Z} \sim \mu^N$ is simply the number of such degenerate states. To generate samples from this ensemble, we use standard Markov-chain Monte Carlo (MC) simulations of the ANNNI chain at zero temperature with single-spin Metropolis updates. Because all domain walls separating the ferromagnetic and period-2 antiphase segments have zero energy cost at $\kappa = 1/2$, the acceptance rate for these updates remains high even at $T = 0$, enabling efficient exploration of the degenerate manifold.

\begin{figure*}
\centering
\includegraphics[width=1.99\columnwidth]{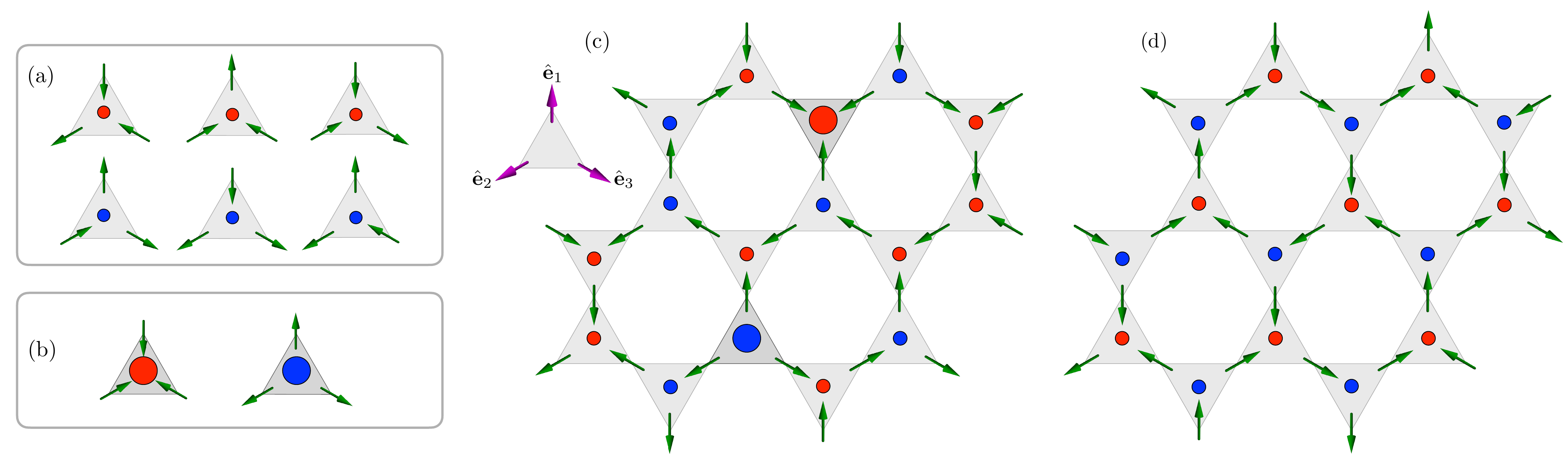}
\caption{(a) Ice-rule configurations (2-in/1-out and 1-in/2-out) for three spins on a triangle unit of the kagome lattice, each carrying an emergent magnetic charge of $Q = \pm 1$ unit. (b) Elementary excitations consist of defect triangles with 3-in or 3-out spin configurations, corresponding to magnetic charges of $Q = \pm 3$. (c) Example of an ice-I state. Small red and blue circles mark triangles that satisfy the ice-rule constraint and carry magnetic charges $Q=+1$ and $-1$, respectively, while large red and blue circles denote defect triangles with $Q=\pm 3$ that violate the constraint. The inset illustrates a single triangular unit cell with the three easy-axis directions $\hat{\mathbf e}_{1,2,3}$ of the kagome lattice. (d) Example of the charge-ordered ice-II phase, in which the $\pm 1$ magnetic charges form a staggered pattern that breaks the $Z_2$ symmetry relating up- and down-triangles. }
    \label{fig:kagome-ice}
\end{figure*}

Spin configurations ${ {\bm \sigma}^{(s)} }$, $s = 1, 2, \cdots, N_{\rm data}$, generated from Monte Carlo simulations are used to train the RBM via the stochastic gradient descent update in Eq.~(\ref{eq:update}). The data-dependent gradient term $\langle \partial E_{\bm \theta} / \partial \bm\theta \rangle_{\rm data}$ is evaluated by clamping the visible spins to the MC-sampled configurations and computing the corresponding hidden-unit expectations from the exact conditional distribution. The model-dependent expectation is approximated using the standard contrastive divergence algorithm (CD-$k$), in which short Gibbs chains initialized from data configurations are evolved for a small number of steps to estimate the model averages. For the relatively simple ANNNI model considered here—where the configuration space is less degenerate and mixing is efficient—CD-$k$ provides a sufficiently accurate and stable estimate of the gradient.

Because time-reversal symmetry remains unbroken at the multiphase point, we fix all visible and hidden biases to zero, leaving the coupling matrix $\bm W$ as the only set of trainable parameters. To monitor the training progress, we track the iteration-to-iteration change in $\bm W$, quantified by
\begin{eqnarray}
\varepsilon_W = \frac{1}{NM} \sum_{i=1}^N \sum_{j = 1}^M \left| W^{(r+1)}_{ij} - W^{(r)}_{ij} \right|.
\end{eqnarray}
Figure~\ref{fig:training-annni} shows $\varepsilon_W$ as a function of the iteration number~$r$, providing a convenient diagnostic of convergence. At early stages, $\varepsilon_W$ exhibits relatively large fluctuations, reflecting the rapid reorganization of the weights as the model adjusts to the statistical structure of the MC-sampled configurations. Within the first few tens of iterations, the amplitude of these fluctuations decreases substantially, indicating that the data-dependent and model-dependent contributions to the gradient are approaching balance. Beyond this transient regime, $\varepsilon_W$ decays smoothly over hundreds of iterations and eventually reaches a small, nearly constant plateau. This saturation signals that the RBM parameters have entered a steady-learning regime in which subsequent updates are minimal, consistent with convergence of the stochastic gradient descent dynamics.


To benchmark the performance of the trained RBM, we consider the spin-spin correlation function defined as
\begin{eqnarray}
	C(r) = \langle \sigma_i \sigma_{i + r} \rangle - \langle \sigma_i \rangle^2
\end{eqnarray}
where the average $\langle \cdots \rangle$ denotes both a spatial average over spin-pairs separated by $r$ within each configuration and an ensemble average over independently sampled configurations. As shown in Fig.~\ref{fig:corr-annni}, the correlation functions computed from RBM-generated samples match those obtained from MC simulations with high fidelity. This agreement demonstrates that the RBM successfully learns not only the uniform probability measure over the degenerate ground-state manifold but also the nontrivial short-range, oscillatory correlations characteristic of the multiphase point $\kappa = 1/2$. In particular, reproducing the oscillatory and exponentially decaying $C(r)$ indicates that the RBM is able to internalize the subtle constraints that define the ground-state manifold---despite the absence of explicit long-range order---highlighting its capability to model constrained, frustration-induced statistical distributions.

\section{Kagome Spin Ice}
\label{sec:kagome}

Next, we apply the RBM framework to learn the correlated, liquid-like phases of kagome spin ice. Kagome spin ice~\cite{wills02} can be viewed as an antiferromagnetic Ising model on the kagome lattice---a two-dimensional network of corner-sharing triangles; see Fig.~\ref{fig:kagome-ice}. In artificial realizations, elongated single-domain ferromagnetic nanomagnets serve as mesoscopic Ising degrees of freedom and are arranged in planar arrays that interact via long-range dipolar fields~\cite{wang06,nisoli13,skjarvo20,chern17,chern21}. In the kagome variant, these nanowires are patterned along the edges of a honeycomb network, the dual of the kagome geometry~\cite{tanaka06,qi08,mengotti10,ladak10,zhang13,anghinolfi2015,canals16,farhan16,yue22,hofhuis22}.

Kagome-ice physics also emerges in several bulk magnets. In canonical pyrochlore spin-ice compounds---Dy$_2$Ti$_2$O$_7$, Ho$_2$Ti$_2$O$_7$---an external field applied along [111] pins one of the four sublattices, producing an effective kagome layer whose remaining spins satisfy kagome-ice rules~\cite{matsuhira02,tabata06,fennell07}. More recently, intrinsic kagome-ice behavior has been identified in the tripod-kagome compound Dy$_3$Mg$_2$Sb$_3$O$_{14}$~\cite{paddison16,dun16,zhao20} and in the quasi-two-dimensional intermetallic magnet HoAgGe~\cite{zhao20}, both hosting frustrated Ising moments on kagome layers. 

In what follows, we focus on the two degenerate ice manifolds characteristic of kagome spin ice. The ice-I phase is defined solely by local ``ice rules" that enforce a two-in/one-out or one-in/two-out configuration on every triangle. The ice-II manifold is further constrained by the emergence of a long-range ordering pattern of emergent magnetic charges, leading to a more strongly correlated yet still liquid-like phase.

\subsection{Ice-I phase}
\label{sec:ice-1}

\begin{figure}
\centering
\includegraphics[width=0.99\columnwidth]{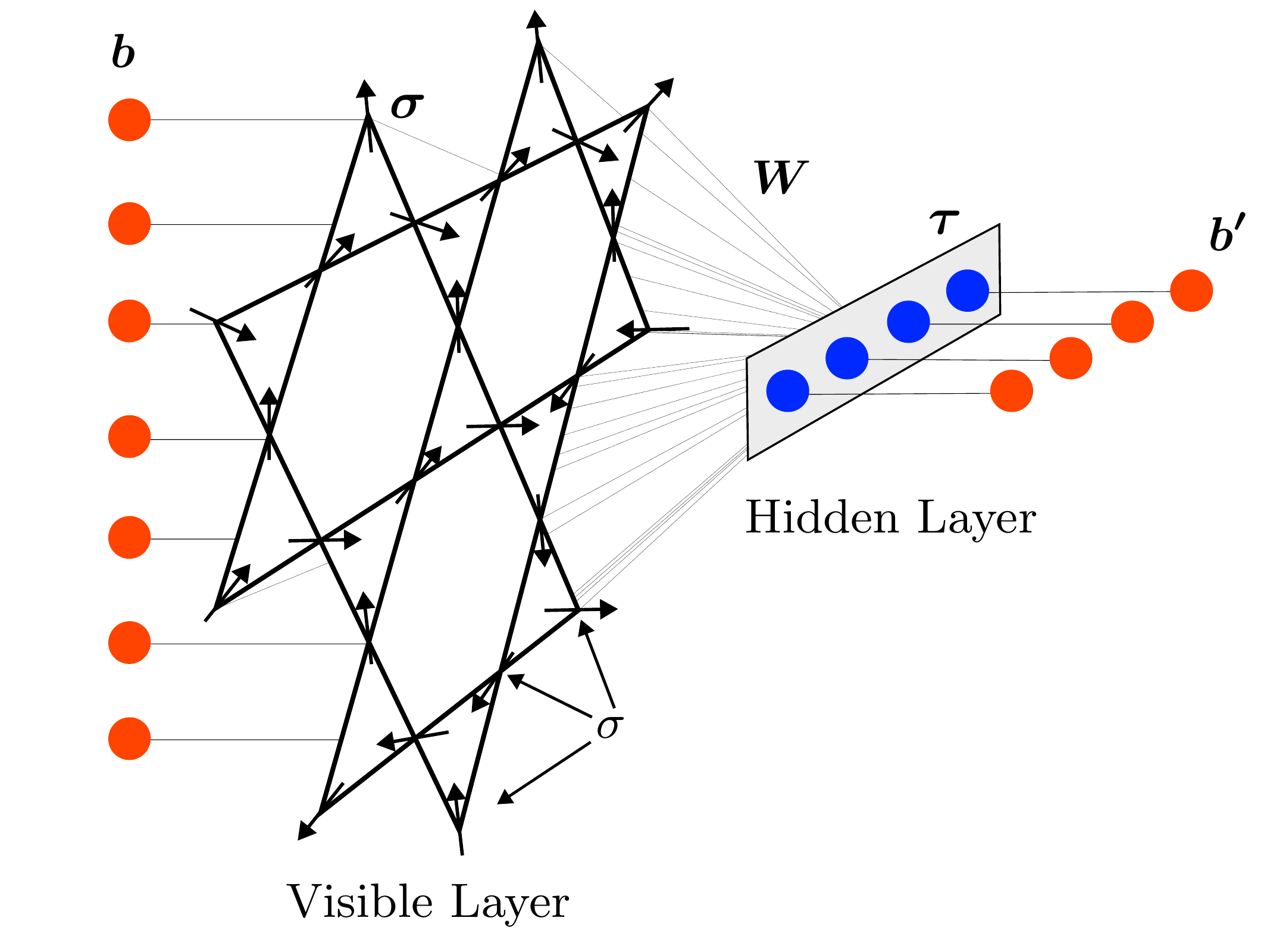}
\caption{Schematic illustration of a RBM for learning the ice phases on the kagome ice. The visible layer represents the Ising spins on the kagome lattice, while the hidden layer encodes latent features associated with the ice-rule constraints and short-range correlations. Trainable parameters include the visible and hidden biases (local fields) and the visible-hidden coupling weights, all of which are optimized using MC-generated spin configurations.}
\label{fig:RBM-kagome}
\end{figure}

To formulate the kagome spin-ice Hamiltonian, we define unit vectors $\hat{\mathbf e}_i$ pointing along the local Ising axes from the center of each up-triangle to its corners. Each magnetic moment is written as $\mathbf S_i = \sigma_i m_0 \hat{\mathbf e}_i$, with $\sigma_i=\pm 1$ denoting its Ising orientation. Because nearest neighbors satisfy $\hat{\mathbf e}_i\cdot \hat{\mathbf e}_j = -1/2$, a ferromagnetic exchange $J_F,\mathbf S_i\cdot\mathbf S_j$ ($J_F<0$) produces an antiferromagnetic Ising coupling,
\begin{eqnarray}
	\label{eq:H-ice1}
	\mathcal{H}_{\mbox{\scriptsize ice-I}} = J \sum_{\langle ij \rangle} \sigma_i \sigma_j = \frac{v}{2}\sum_{\alpha} Q_\alpha^2 + E_0,
\end{eqnarray}
with $J=|J_F|/2$. The second form rewrites the Hamiltonian in terms of magnetic charges $Q_\alpha$ on triangles.
A convenient picture is the dumbbell model~\cite{castelnovo08}, where each dipole is replaced by a bar magnet of length $\ell$ whose poles of charge $q_m=m_0/\ell$ sit at neighboring triangle centers. The net charge on triangle $\alpha$ is
\begin{eqnarray}
	Q_\alpha = \pm q_m \sum_{i \in \alpha} \sigma_i, 
\end{eqnarray}
with the sign distinguishing up- and down-triangles. In these units, $v=J\ell^2/m_0^2$.
Minimizing Eq.~(\ref{eq:H-ice1}) requires each triangle to carry the smallest possible charge, $Q_\alpha=\pm 1$, corresponding to 2-in/1-out or 1-in/2-out spin arrangements, as shown in Fig.~\ref{fig:kagome-ice}(a). These are the kagome-ice analog of the Bernal-Fowler ice rules~\cite{bernal33}.   The correlated states defined by these 2-in/1-out or 1-in/2-out constraints constitute the ice-I manifold. Their exponentially large number yields a finite residual entropy per spin, $S_0 \approx \frac{1}{3}\ln(9/2)$~\cite{wills02}.  The magnetic-charge representation in Eq.~(\ref{eq:H-ice1}) also makes evident that excitations out of this manifold correspond to defect triangles with $Q = +3$ or $-3$, i.e., 3-in or 3-out configurations; see Fig.~\ref{fig:kagome-ice}(b). These defects act as magnetic monopoles within the kagome-ice framework and control the low-energy excitations above the ice-rule manifold. A representative spin-disordered ice-I state, including an example of a $Q=\pm 3$ defect pair, is shown in Fig.~\ref{fig:kagome-ice}(c).

Here we implement an RBM to learn the ice-I manifold of kagome spin ice; a kagome-specific schematic is shown in Fig.~\ref{fig:RBM-kagome}. The visible units of the RBM encode the Ising spins on the kagome lattice in a fixed ordering, while effective interactions among these spins are mediated through their couplings to the hidden layer, as reflected in the effective energy expression Eq.~(\ref{eq:E_eff}). As discussed in Sec.~\ref{sec:RBM}, training adjusts the RBM parameters so that the marginal distribution Eq.~(\ref{eq:p_marginal}) provides an accurate approximation to the zero-temperature ground-state manifold of Eq.~(\ref{eq:H-ice1}), i.e. $\pi_{\bm\theta}(\bm\sigma) \approx \pi_{\scriptsize\mbox{ice-I}}(\bm\sigma; \, T = 0)$, which is a uniform distribution over all configurations satisfying the local ice rules. Because nonzero bias fields $\bm b$ and $\bm b'$ explicitly break time-reversal symmetry, $\bm\sigma \to -\bm\sigma$, and the ice-I manifold is itself time-reversal invariant, we set both biases to zero throughout training.

\begin{figure}
\centering
\includegraphics[width=0.99\columnwidth]{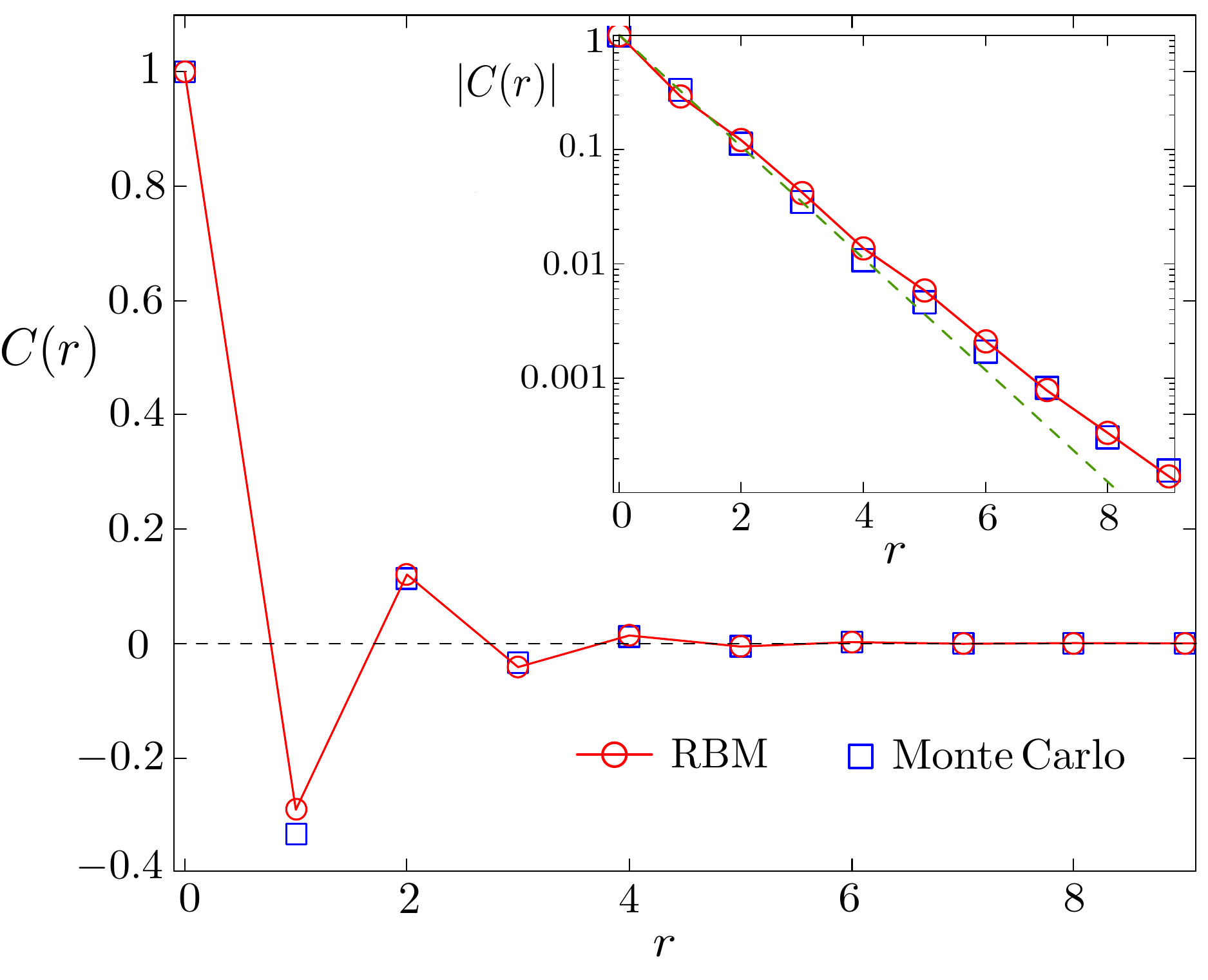}
\caption{Comparison of the spin-spin correlation function $C(r)$ obtained from MC sampling and from RBM-generated configurations for the $T = 0$ ice-I manifold of $\mathcal{H}_{\scriptsize \mbox{ice-I}}$. The inset shows $|C(r)|$ on a semi-logarithmic scale, highlighting the short-range nature of the correlations. The dashed line indicates an exponential fit, $\exp(-r/\ell)$, with correlation length $\ell \approx 0.89$.}
\label{fig:corr-ice1}
\end{figure}

The training data for the kagome ice-I phase were generated using standard Markov-chain Monte Carlo (MC) simulations on a $32 \times 32$ lattice ($N = 3 \times 32^2$ spins) at $T = 0$. Unlike the more constrained pyrochlore ice or the charge-ordered ice-II manifold discussed below, kagome ice-I possesses an extensive manifold of local zero modes: many single-spin flips preserve the 2-in/1-out (or 1-in/2-out) constraint on every triangle. Consequently, simple single-spin Metropolis updates efficiently explore the manifold without becoming confined to restricted sectors of configuration space.

For the RBM training, however, accurate sampling of the model distribution is essential in order to avoid systematic bias in the gradient estimation. As discussed in more detail in Appendix~\ref{app:numerical-details}, we therefore employ persistent contrastive divergence (PCD) rather than the standard CD-$k$ algorithm. In large and highly degenerate systems such as kagome ice, CD-$k$—which repeatedly initializes short Gibbs chains from data configurations—can produce biased estimates of the model expectation values, potentially leading to spurious symmetry breaking or imbalance between degenerate sectors of the manifold. By contrast, PCD maintains a set of persistent Markov chains throughout training, allowing them to track the slowly evolving stationary distribution of the RBM. This substantially reduces gradient bias and ensures that the learned model faithfully represents the uniform $T=0$ ice manifold without artificial sector selection.

Figure~\ref{fig:corr-ice1} shows that the RBM trained on these MC samples quantitatively reproduces the spin–spin correlation function. The correlations display the characteristic short-range structure imposed by the ice rules: an oscillatory component with a period of one lattice spacing, originating from the nearest-neighbor antiferromagnetic interaction, modulated by a strong exponential envelope $e^{-r/\ell}$. The exponential decay, clearly visible in the semi-logarithmic inset, reflects geometric frustration that suppresses long-range order and yields a correlation length $\ell \sim \mathcal{O}(1)$ lattice spacing. Small deviations at large distances arise from finite-size effects. The close agreement between RBM-generated configurations (red circles) and MC data (blue triangles) across all distances demonstrates that the trained RBM faithfully captures both the local ice-rule constraints and the short-range correlated nature of the kagome ice-I manifold.

\subsection{Ice-II phase}

\label{sec:ice-2}

In real kagome-ice materials, long-range dipolar interactions play a central role. In artificial kagome ices in particular, the nearest-neighbor coupling $J$ that stabilizes the ice-I phase arises entirely from the dipole-dipole interaction between magnetic nano-islands. While the nearest-neighbor model supports a massively degenerate ice-I manifold, dipolar interactions lift this degeneracy and generate additional correlations. At low temperature they eventually drive true long-range order, but at intermediate temperatures they stabilize an emergent ice-II phase: a partially ordered state in which the residual charges $Q_\alpha=\pm 1$ arrange in an alternating pattern on the dual honeycomb lattice.

This charge ordering is most naturally interpreted through the effective Coulomb interaction between triangle charges, $(\mu_0 q_m^2 / 8\pi)\, Q_\alpha Q_\beta / |\mathbf r_\alpha - \mathbf r_\beta|$, which gives the leading-order approximation of the dipolar interaction. To capture the essential physics of the ice-II phase within a minimal theoretical framework, we consider a spin-ice Hamiltonian supplemented by a short-range interaction between neighboring triangle charges:
\begin{eqnarray}
	\label{eq:H-ice2}
	\mathcal{H}_{\mbox{\scriptsize ice-II}}  = \frac{v}{2}\sum_{\alpha} Q_\alpha^2 + K \sum_{\alpha\beta} Q_\alpha Q_\beta,
\end{eqnarray}
where Greek indices label the triangular plaquettes of the kagome lattice. The coefficient $v = J \ell^2 / m_0^2$ represents the local self-energy of magnetic charges generated by the nearest-neighbor antiferromagnetic exchange $J$. Minimizing this term enforces $Q_\alpha = \pm 1$, recovering the 2-in/1-out (or 1-in/2-out) ice rule discussed in Sec.~\ref{sec:ice-1}. The second term, proportional to $K > 0$, introduces an antiferromagnetic interaction between charges on adjacent triangles. This repulsive coupling selects a staggered pattern of $Q_\alpha = \pm 1$ on the honeycomb dual lattice, thereby breaking the $Z_2$ symmetry: all up-triangles acquire the same charge (say $Q=+1$), while all down-triangles carry the opposite charge. This simple extension thus encapsulates the emergence of charge order characteristic of the kagome ice-II phase.

Importantly, the spins themselves remain disordered within the charge-ordered state; see Fig.~\ref{fig:kagome-ice}(c) for an example. The residual degeneracy can be quantified by mapping every charge-ordered configuration to a dimer covering of the honeycomb lattice: in each triangle with $Q_\alpha=\pm 1$, two majority spins point into (or out of) the triangle, and the minority spin points in the opposite direction. Associating these minority spins with dimers on the dual honeycomb lattice establishes a one-to-one correspondence with perfect dimer coverings. Because the number of honeycomb dimer coverings grows exponentially with system size, the ice-II phase retains a finite residual entropy, $S = 0.108$ per spin~\cite{udagawa02}. The ice-II phase therefore exemplifies ordering driven by emergent degrees of freedom—magnetic charges—rather than by the microscopic spins themselves, and highlights the hierarchical structure of correlations in kagome spin ice.

\begin{figure}
\centering
\includegraphics[width=0.99\columnwidth]{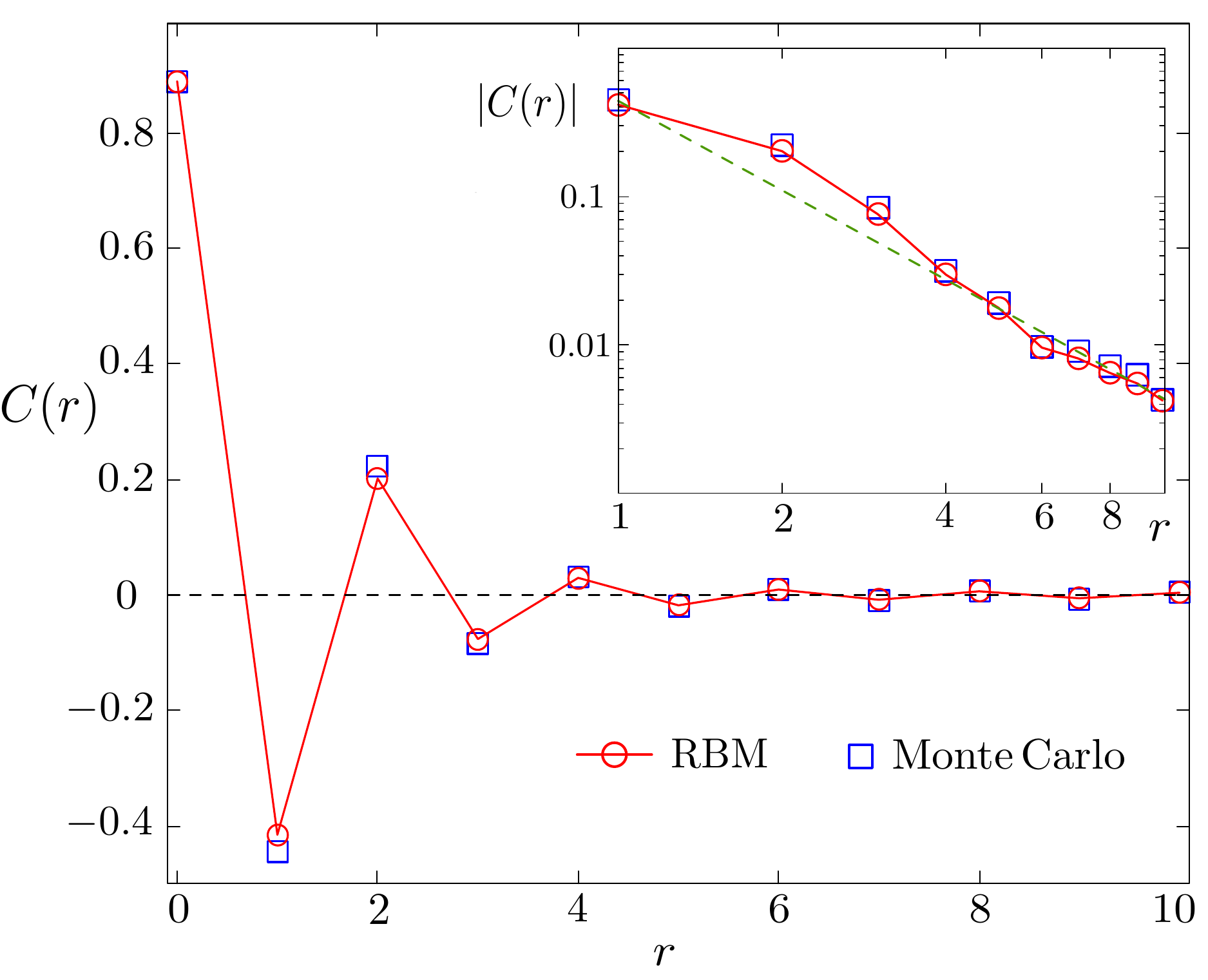}
\caption{Comparison of the spin–spin correlation function $C(r)$ for the kagome ice-II phase at $T=0$ with nearest-neighbor charge repulsion $K=0.1J$. Results from Markov-chain Monte Carlo simulations are compared with correlations computed from RBM-generated configurations, demonstrating excellent agreement across all distances. The inset shows the absolute value $|C(r)|$ on a log–log scale; the dashed line indicates a $1/r^{2}$ power-law decay, highlighting the algebraic nature of correlations in the charge-ordered ice-II phase.}
\label{fig:corr-ice2}
\end{figure}

To train RBMs for the ice-II phase, we perform Monte Carlo simulations of the Hamiltonian in Eq.~(\ref{eq:H-ice2}) at zero temperature, where the short-range charge interaction stabilizes a staggered charge-ordered state. Throughout, the nearest-neighbor charge repulsion is fixed at $K=0.1J$. The emergence of charge order imposes constraints beyond the conventional ice rules, substantially reducing the accessible configuration space and rendering the ground-state manifold far more restricted than in the ice-I case. In this charge-ordered background, a single-spin flip generically violates the local $Q_\alpha=\pm 1$ condition, either creating high-energy $Q=\pm 3$ monopole defects or producing misplaced $Q=\pm 1$ charges that disrupt the staggered arrangement. At $T=0$, where any defect carries a finite energy penalty, such local moves are strictly forbidden, and naive single-spin Metropolis updates become entirely ineffective.

The Monte Carlo sampling of the ice-II manifold at zero temperature is therefore carried out exclusively using nonlocal loop updates. These updates construct closed loops of spins whose collective reversal preserves the $Q_\alpha=\pm 1$ constraint on every triangle while maintaining the global staggered charge pattern. Because the net charge on each triangle remains unchanged, flipping an entire loop connects two distinct configurations within the same charge-ordered manifold without generating monopoles or other defects. This loop algorithm ensures ergodic exploration of the restricted ground-state manifold while exactly respecting the zero-energy constraints imposed by the Hamiltonian.

The onset of charge order also breaks a $Z_2$ symmetry associated with time-reversal. Consequently, the effective RBM Hamiltonian in Eq.~(\ref{eq:E_eff}) must be capable of explicitly breaking time-reversal symmetry, which underscores the necessity of including local bias fields in the model. In contrast to the ice-I case—where symmetry considerations allow the biases to be fixed to zero—RBMs trained without local fields fail to accurately represent ice-II configurations.
Accordingly, in the RBM training for the ice-II phase, we treat $\bm b$, $\bm b'$, and $\bm W$ as fully trainable parameters. These parameters are optimized such that the marginal distribution in Eq.~(\ref{eq:p_marginal}) faithfully approximates the degenerate $T=0$ manifold, i.e., $\pi_{\bm\theta}(\bm\sigma) \approx \pi_{\scriptsize\mbox{ice-II}}(\bm\sigma; \, T = 0)$, which is again a uniform distribution over all spin configurations satisfying the restricted ice rules: $\pi_{\scriptsize\mbox{ice-II}}(\bm\sigma; \, T = 0) = 1/\mathcal{Z}$ Here $\mathcal{Z}=\exp(N S_0)$ denotes the ground-state degeneracy, with residual entropy per spin $S_0 \approx 0.108$.

Fig.~\ref{fig:corr-ice2} compares the spin-spin correlation function $C(r)$ of the kagome ice-II phase obtained from Markov-chain MC simulations with that generated by a trained RBM. The presence of the staggered charge order enhances the short-range antiferromagnetic correlations between spins. the RBM-sampled values (red circles) track the Monte Carlo results (blue triangles) semiquantitatively, capturing the oscillatory behavior and the rapid suppression of correlations that follow from the disordered spin background. The overall agreement demonstrates that, once equipped with symmetry-breaking bias fields, the RBM successfully learns both the local and intermediate-range structure of the ice-II phase.

The inset of Fig.~\ref{fig:corr-ice2} presents $|C(r)|$ on a log-log scale, revealing a clear power-law envelope. The dashed line corresponds to a $1/r^2$ power-law decay, in agreement with the expected asymptotic form for correlations in the ice-II manifold, where long-range charge order coexists with algebraically decaying spin correlations inherited from an underlying dimer-like description~\cite{moessner03}. That the RBM reproduces this power-law behavior---despite the absence of explicit long-range couplings in its architecture---indicates that the hidden layer has effectively learned the emergent constraints imposed by the staggered charge pattern. 

\begin{figure}
\centering
\includegraphics[width=0.99\columnwidth]{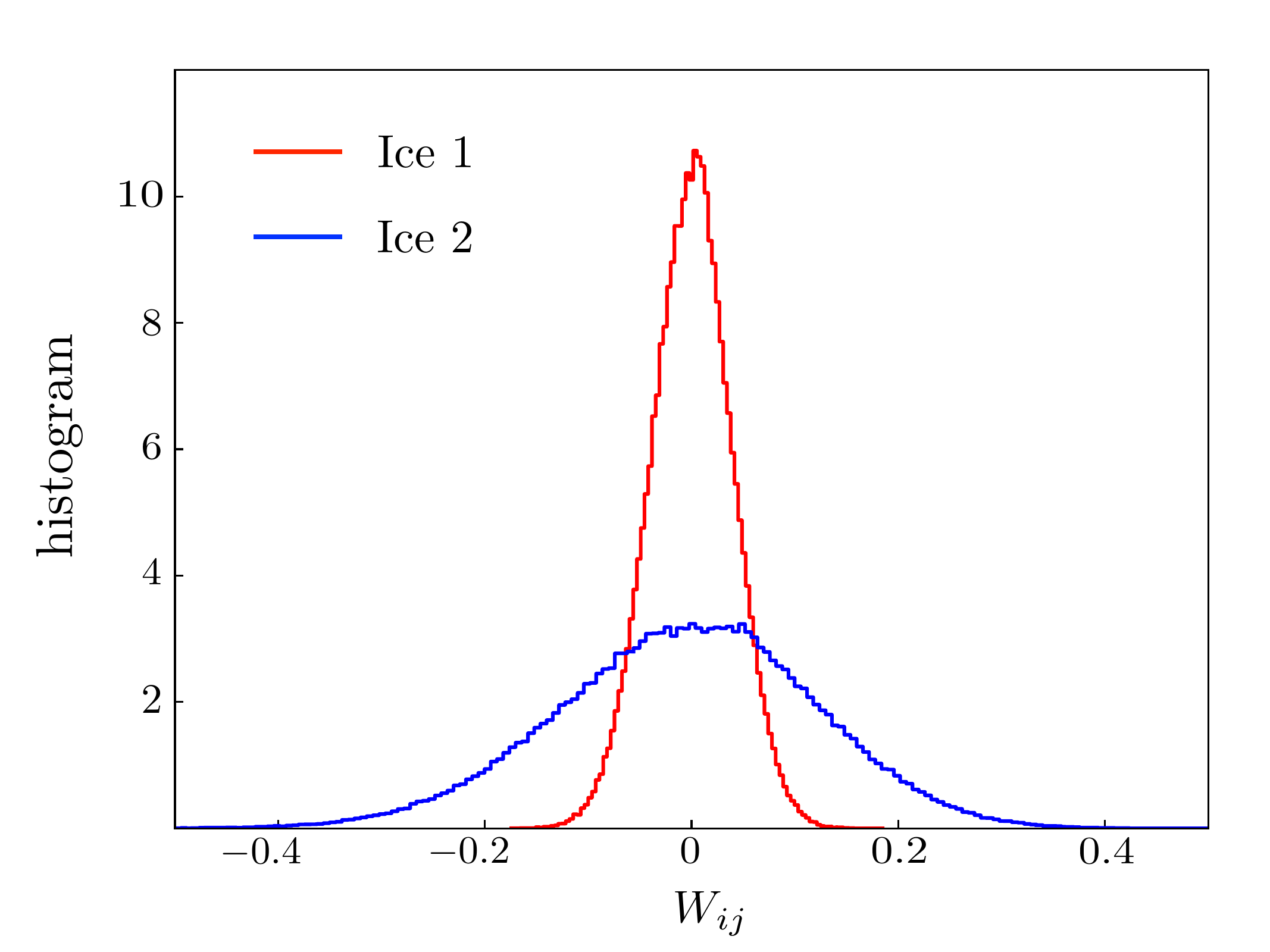}
\caption{Histogram of the learned RBM couplings $W_{ij}$ for the kagome ice-I and ice-II phases. The ice-I distribution (blue) is relatively narrow and centered near zero, consistent with the highly degenerate and weakly constrained ice-I manifold. By contrast, the ice-II distribution (red) is noticeably broader, with larger-magnitude weights. This increased variance reflects the stronger constraints and reduced entropy of the charge-ordered ice-II phase.}
\label{fig:hist-W}
\end{figure}

The broader distribution of the learned couplings $\bm W$ in the ice-II phase can be understood as a consequence of its more strongly constrained and lower-entropy ground-state manifold. Compared with ice-I, which possesses an extensive set of local zero modes and a relatively flat, highly degenerate configuration space, the charge-ordered ice-II phase occupies a smaller and more structured subset of spin configurations. In particular, the additional charge-order constraint reduces the residual entropy and imposes more rigid correlations between spins. As shown in Fig.~\ref{fig:hist-W}, the histogram of $W_{ij}$ for ice-II exhibits a visibly larger variance and heavier tails than that of ice-I, indicating that the RBM develops stronger visible–hidden couplings in order to encode these enhanced constraints. To faithfully represent the restricted ice-II manifold, the model must more sharply suppress configurations that violate the charge-order condition while enhancing the weight of symmetry-compatible states, which naturally leads to larger-magnitude parameters.

From the perspective of the effective RBM energy landscape, ice-I corresponds to a comparatively shallow and nearly isotropic manifold, reflected in the narrower and more concentrated weight distribution seen in Fig.~\ref{fig:hist-W}. By contrast, the ice-II phase exhibits more pronounced energy basins associated with broken time-reversal symmetry and emergent sublattice order. Capturing these sharper distinctions requires stronger correlations mediated by the hidden units, resulting in the broader spread of $W_{ij}$ observed in the figure. In this sense, the enhanced variance of the weight distribution for ice-II directly mirrors its reduced entropy and increased structural coherence, and thus provides an informative diagnostic of how the RBM internalizes the underlying physical constraints.

\begin{figure}
\centering
\includegraphics[width=0.99\columnwidth]{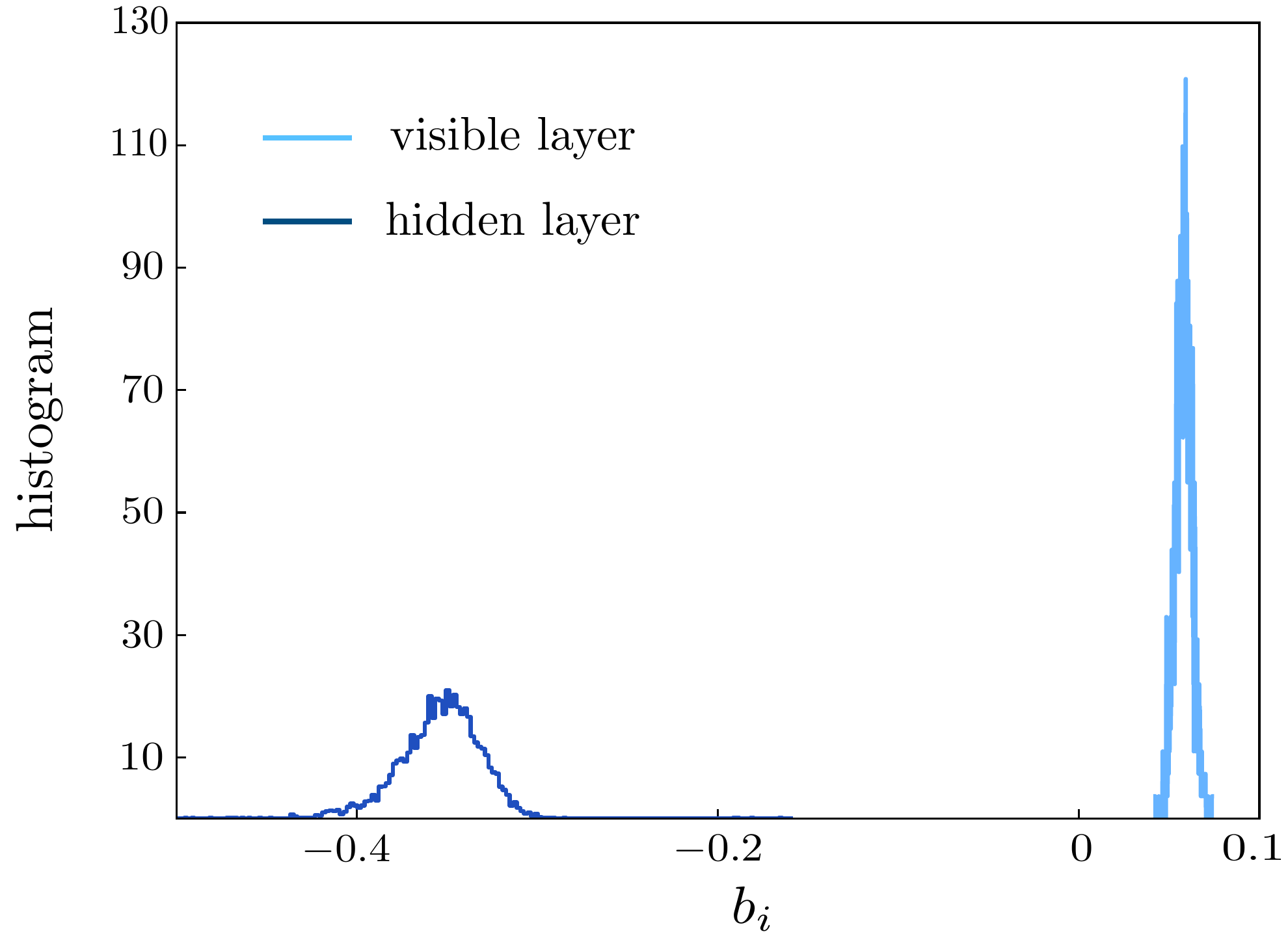}
\caption{Histograms of the RBM local fields for the ice-II phase. Light curves correspond to the visible-layer biases, while dark curves denote the hidden-layer biases. The visible-layer fields predominantly cluster around one positive sign, whereas the hidden-layer fields peak around the opposite sign. This clear sign asymmetry reflects the spontaneous breaking of time-reversal symmetry associated with the charge-ordered ice-II phase. }
\label{fig:hist-b}
\end{figure}

The distributions of the local bias fields learned by the RBM provide a direct probe of how the model captures the broken symmetries and correlation structure of the ice-II phase. As shown in Fig.~\ref{fig:hist-b}, the visible-layer fields $\bm b$ (light curves) are consistently concentrated on the same sign, whereas the hidden-layer fields $\bm b'$ form a distribution centered at the opposite sign.  The emergence of uniformly signed visible-layer fields is particularly significant. In the RBM effective energy, these fields act as local “magnetic fields’’ that bias the visible spins toward one orientation. A uniform sign in the visible-layer fields acts as an effective global bias on the physical spins and directly reflects the broken time-reversal $Z_2$ symmetry of the ice-II phase. Once monopole charges order, the local spin environments acquire a preferred orientation, and the RBM encodes this by assigning visible biases of consistent sign.
The hidden-layer fields show the complementary structure: they peak around the opposite sign, indicating that the hidden units activate in a way that reinforces the broken-symmetry pattern. This opposing-sign structure in the two layers demonstrates that the RBM internalizes the essential symmetry breaking of ice-II and represents it through a coherent organization of its bias fields.

\section{Conclusion and Outlook}

In this work we have demonstrated that Restricted Boltzmann Machines (RBMs) offer a powerful and flexible generative framework for learning the statistical structure of frustrated magnetic systems. As a controlled benchmark, we first applied the RBM to the one-dimensional ANNNI model at its multiphase point, where the ground state is highly degenerate and characterized by oscillatory, exponentially decaying spin-spin correlations. The trained RBM successfully reproduced these features with quantitative accuracy, showing that even a shallow generative model is capable of capturing nontrivial emergent correlations arising from competing interactions.

Building on this foundation, we employed RBMs to learn the spin configurations of kagome spin ice, whose ice-I manifold exhibits extensive degeneracy and short-range correlations enforced by local 2-in/1-out constraints. Using Monte Carlo samples as training data, the RBM faithfully captured these ice rules and accurately reproduced the spin-spin correlation functions of the ice-I phase. We further examined the more intricate ice-II phase, where long-range charge order emerges from effective Coulomb interactions between magnetic charges. Here, we found that correct modeling requires RBMs with nonzero, uniform-sign bias fields, reflecting the underlying breaking of time-reversal symmetry. Together, these results demonstrate that RBMs are capable of learning both local constraints and emergent correlations in frustrated magnets, even when the underlying manifolds are highly degenerate or partially ordered.

Our findings highlight several promising directions for future work. Extending the present approach to larger system sizes and deeper or more expressive architectures---such as convolutional RBMs and tensor-network inspired deep Boltzmann machines---may further enhance the model's ability to capture long-range emergent structures in more complex frustrated systems. 
Finally, the success of RBMs in learning kagome spin-ice manifolds suggests that generative models may serve as useful tools for probing otherwise intractable frustrated systems, including classical spin liquids, Coulomb-phase systems, and systems with gauge constraints. By providing compact probabilistic representations of these highly correlated states, generative models hold the potential to complement conventional simulation methods and offer new perspectives on emergent phenomena in frustrated magnetism.

\label{sec:conclusion}

\begin{acknowledgments}
This work was supported by the US Department of Energy Basic Energy Sciences under Award No. DE-SC0020330. The authors also acknowledge the support of Research Computing at the University of Virginia.
\end{acknowledgments}

\appendix
\section{Numerical Implementation Details}

\label{app:numerical-details}


As discussed in Sec.~\ref{sec:RBM}, training the RBM amounts to learning a parametric probability distribution $\pi_{\boldsymbol{\theta}}(\bm{\sigma})$ that approximates the physical Boltzmann distribution $\pi_{\mathrm{phys}}(\bm{\sigma})$ defined in Eq.~(\ref{eq:phys_distribution}). Here $\bm{\sigma}=\{\sigma_1,\dots,\sigma_N\}$ denotes a visible-layer spin configuration. The discrepancy between the two distributions is quantified by the Kullback–Leibler (KL) divergence,\begin{equation}
    D_{\mathrm{KL}}\!\left(\pi_{\mathrm{phys}} \,\middle\|\, \pi_{\boldsymbol{\theta}} \right) = \sum_{\boldsymbol{\sigma}} \pi_{\mathrm{phys}}(\boldsymbol{\sigma}) 
    \ln\frac{ \pi_{\mathrm{phys}}(\boldsymbol{\sigma})} {\pi_{\boldsymbol{\theta}}(\boldsymbol{\sigma}) }.
\label{eq:KL}
\end{equation}
which measures how well the RBM reproduces the statistical weights of physical configurations. Minimizing $D_{\mathrm{KL}}$ with respect to the model parameters $\boldsymbol{\theta}=\{\bm{W},\bm{b},\bm{b}' \}$ is equivalent to maximizing the likelihood of the training data under the RBM ansatz and constitutes the central optimization task. The parameters are initialized as $\boldsymbol{\theta}^{(0)}=\{\bm{W}^{(0)},\bm{b}^{(0)},\bm{b}^{\prime(0)} \}$, with vanishing biases $\bm{b}^{(0)}=\bm{b}^{\prime(0)}=0$ and weights $W_{ij}^{(0)}$ drawn independently from a normal distribution with zero mean and standard deviation $0.01$. Such small random initialization places the model near a high-entropy, weakly correlated state, allowing nontrivial correlations and symmetry breaking to emerge gradually during training.

Parameter updates follow stochastic gradient descent,
\begin{equation}
    \Delta \boldsymbol{\theta}^{(n)} = \boldsymbol{\theta}^{(n+1)} - \boldsymbol{\theta}^{(n)} = - \eta \, \nabla_{\boldsymbol{\theta}} D_{\mathrm{KL}},
\label{eq:sgd_update}
\end{equation}
where $\eta$ denotes the learning rate. Differentiating the KL divergence yields, for example, for the weight matrix,
\begin{equation}
\frac{\partial D_{\mathrm{KL}}}{\partial W_{ij}}
=
- \bigl\langle
\sigma_i \tau_j
\bigr\rangle_{\pi_{\mathrm{phys}}(\boldsymbol{\tau} \mid \boldsymbol{\sigma})}
+
\bigl\langle
\sigma_i \tau_j
\bigr\rangle_{\pi_{\boldsymbol{\theta}}(\boldsymbol{\sigma}, \boldsymbol{\tau})} .
\label{eq:rbm_grad}
\end{equation}
This expression has a clear physical interpretation. The first term (“data average”) measures correlations between visible and hidden units when the visible configuration is clamped to a Monte Carlo sample drawn from the physical distribution. Because the RBM factorizes conditionally, the hidden expectation values can be computed exactly from $\pi_{\boldsymbol{\theta}}(\bm{\tau}\mid\bm{\sigma})$.
The second term (“model average”) requires sampling from the full joint model distribution $\pi_{\boldsymbol{\theta}}(\bm{\sigma},\bm{\tau})$. This step is computationally challenging because evaluating exact expectations would require summing over an exponentially large configuration space through the partition function $Z_{\boldsymbol{\theta}}$. In practice, one therefore relies on Markov-chain Monte Carlo sampling within the RBM itself.

Because exact evaluation would require computing the partition function $Z_{\boldsymbol{\theta}}$ over an exponentially large configuration space, one resorts to Markov-chain Monte Carlo sampling within the RBM. A natural choice is Gibbs sampling, which alternately updates visible and hidden variables according to their exact conditional distributions:
\begin{equation*}
	\boldsymbol{\sigma}^{(0)} \to \boldsymbol{\tau}^{(0)} \to \boldsymbol{\sigma}^{(1)} \to \boldsymbol{\tau}^{(1)} \to \cdots \to \boldsymbol{\sigma}^{(k)} \to \boldsymbol{\tau}^{(k)} ,
\end{equation*}
where each transition is performed using the conditional probabilities $\pi_{\boldsymbol{\theta}}(\boldsymbol{\tau} \mid \boldsymbol{\sigma})$ and $\pi_{\boldsymbol{\theta}}(\boldsymbol{\sigma} \mid \boldsymbol{\tau})$. Because each conditional update satisfies detailed balance, this Markov chain converges, in the limit of long sampling time, to the stationary Boltzmann distribution of the RBM. In principle, one could run this chain to equilibrium at every parameter update and use the resulting samples to estimate the model average in Eq.~(\ref{eq:rbm_grad}). However, such full equilibration at every training step is computationally prohibitive for the large spin systems considered here.

In principle, one could equilibrate the Gibbs chain at each parameter update to obtain unbiased estimates of the model expectation. In practice, however, full equilibration is computationally prohibitive for large systems such as the kagome lattices studied here. Contrastive divergence (CD-$k$) addresses this issue by initializing the Gibbs chain from a configuration drawn from the training dataset and performing only a small number $k$ of alternating updates before estimating the model expectation. While CD-$k$ often performs surprisingly well for moderate system sizes, it provides a biased gradient estimate because the truncated chain samples from a distribution that remains closer to the data distribution than to the true model distribution~\cite{Fischer2014}.

This bias becomes particularly pronounced in systems with large configuration spaces, slow mixing, or highly degenerate manifolds. In the kagome ice-I phase, for example, the ground-state manifold consists of symmetry-related sectors with equal statistical weight. Physical observables such as the unit-cell spin sum $Q_\alpha={-1,1}$ must therefore exhibit symmetric distributions. During CD-$k$ training, however, we observe that the model gradually develops a preference for one sector. This spurious symmetry breaking originates from insufficient equilibration of the Gibbs chain: because each chain is repeatedly reinitialized from the data and evolved only briefly, it fails to sample the full model distribution and instead drifts toward metastable regions of configuration space. The resulting gradient estimate is systematically biased.

To reduce this bias, we employ persistent contrastive divergence (PCD)~\cite{Tielman08}. The essential modification relative to CD-$k$ is that the Markov chains used to estimate the model expectation are not reinitialized from the data at every iteration. Instead, a set of persistent chains is maintained throughout the entire training process. At each parameter update, these chains are advanced by a small number of Gibbs steps, and their current states are used to estimate the model expectation. The conceptual advantage of PCD relies on the fact that the model parameters evolve gradually during training, provided the learning rate is sufficiently small. Consequently, the stationary distribution $\pi_{\boldsymbol{\theta}}$ changes only adiabatically between successive updates. A persistent Markov chain that was approximately equilibrated with respect to $\pi_{\boldsymbol{\theta}^{(n)}}$ remains close to equilibrium for $\pi_{\boldsymbol{\theta}^{(n+1)}}$, requiring only a few additional Gibbs steps to re-equilibrate. In this sense, PCD tracks the slowly drifting Boltzmann distribution of the RBM, rather than repeatedly restarting the sampling process from the data manifold.

From a statistical perspective, CD-$k$ estimates the model expectation using a distribution obtained after $k$ applications of the Gibbs transition operator to the data distribution, whereas PCD estimates it using a distribution that remains close to the stationary model distribution throughout training. In the limit of vanishing learning rate and sufficiently long Gibbs chains, PCD approaches exact maximum-likelihood learning. In practice, even a small number of Gibbs steps per update significantly reduces the systematic bias compared to CD-$k$.
In our simulations of kagome spin ice, PCD eliminates the artificial imbalance between symmetry-related sectors observed under CD-$k$ and restores the expected symmetry of observables within the degenerate manifold. All results presented for both ice-I and ice-II phases are therefore obtained using the PCD algorithm.

To assess the RBM’s capacity to represent these extensively degenerate manifolds, we consider systems of size $N=3\times32^2$ (ice-I) and $N=3\times24^2$ (ice-II). The training dataset consists of 5000 independent Monte Carlo configurations, generated with sufficient thinning to suppress autocorrelation effects. We employ a hidden layer with $N_h=5000$ units, providing substantial representational capacity while maintaining stable optimization.
For the ice-I phase, the visible and hidden biases $\bm{b}$ and $\bm{b}'$ are fixed to zero in order to preserve the time-reversal symmetry of the Hamiltonian. In contrast, nonzero local fields are allowed for ice-II as time-reversal symmetry is broken in the presence of magnetic charge order. Training is performed for 2000 epochs using the Adam optimizer, with a scheduled learning rate: $\eta=10^{-4}$ for the first 1000 epochs and $\eta=10^{-5}$ for the remaining 1000 epochs. No weight decay is applied. All computations are carried out in PyTorch with GPU acceleration on an NVIDIA RTX A6000.

\bibliography{ref}

\end{document}